\documentclass{PoS}
\title{Charm and bottom heavy baryon mass spectrum from lattice QCD with 2+1 flavors}

\ShortTitle{Heavy baryon mass spectrum}

\author{\speaker{Heechang Na} and Steven Gottlieb\\
        Department of Physics, Indiana University, Bloomington, Indiana 47405, USA\\
        E-mail: \email{heena@indiana.edu}, \email{sg@indiana.edu}}

\abstract{We present results for the mass spectrum of charm and bottom heavy baryons, using MILC coarse lattice configurations with 2+1 flavors. Clover heavy quark propagators with the Fermilab interpretation and improved staggered light quark propagators are used to construct two point functions with local operators for different flavor and spin states.}

\FullConference{The XXV International Symposium on Lattice Field Theory\\
		 July 30-4 August 2007\\
		 Regensburg, Germany}

\begin{document}

\section{Introduction}
Heavy baryons have been extensively investigated by experimental and theoretical approaches.
From experiment, the singly charmed heavy baryon 
mass spectrum is well known; however, the other heavy baryon masses are 
only crudely known.  Recently, from the D$\emptyset$ \cite{D0}  and CDF \cite{CDF} collaborations, there are measurements of the mass of the singly bottom baryon $\Xi_b^-$. 
From lattice QCD, there are several quenched 
calculations for the heavy baryon mass spectrum [3--8],
and most results are in fair agreement with observed values.
In this project, we apply 
lattice QCD with dynamical sea quarks for the same objective.  

Previously, we have presented the preliminary results for the mass spectrum of singly charmed heavy baryons \cite{na}. Here, we present the singly charmed and bottom heavy baryon as well as the doubly charmed and bottom heavy baryon mass spectrum. 
In this work, we use two different interpolating operators studied in 
Ref.~\cite{UKQCD}, and construct two-point functions using the 
method of Wingate \emph{et al}. \cite{wingate} to combine staggered 
propagators for the light valence quarks
and a Wilson type (clover) propagator for the heavy valence quark.
In addition, we show that there is no taste mixing problem by studying the explicit spin and taste structure of operators.

\section{Construction of two-point functions}
The interpolating operators 
to describe singly heavy baryons are 
\cite{UKQCD} 
\begin{equation}
\mathcal{O}_5 = \epsilon_{abc} ( \psi_1^{aT} C \gamma_5 \psi_2^b ) \Psi_H^c , \;\;\;\;\;\;\;
\mathcal{O}_\mu = \epsilon_{abc} ( \psi_1^{aT} C \gamma_\mu \psi_2^b ) \Psi_H^c,
\label{operator}
\end{equation}
where $\epsilon_{abc}$ is the Levi-Civita tensor, $\psi_1$
and $\psi_2$ are light valence quark fields for up, down, or strange quarks,
$\Psi_H$ is the heavy valence quark field for the charm or the bottom quark,
 $C$ is the charge conjugation matrix, and $a$, $b$, and $c$ are color indices. 
Basically, $\mathcal{O}_5$ is the operator for 
$s^\pi = 0^+$, and  $\mathcal{O}_\mu$ is for
$s^\pi = 1^+$, where $s^\pi$ is the spin parity state for the light quark pair.  
As can be seen, the spinor indices of the light quark fields
are contracted together, so the spinor index of the operator comes from 
$\Psi_H$.
Furthermore, each heavy baryon is obtained by choosing appropriate quark flavor combinations, and, for doubly heavy baryons, we can simply interchange the heavy quark field and the light quark fields in Eq.~\ref{operator} \cite{woloshyn1}.

Since we use staggered fermion propagators for the light quarks, and a Wilson type propagator for
the heavy quark, 
we will use the method of Wingate \emph{et al}. \cite{wingate} to convert to a naive quark propagator starting from a staggered 
propagator. They show
\begin{equation}
G_\psi(x,y)=\Omega(x)\Omega^\dagger(y)G_\chi(x,y),
\label{win1}
\end{equation}
where
\begin{equation}
\Omega(x)=\prod_\mu(\gamma_\mu)^{x_\mu/a},
\label{omega}
\end{equation}
$G_\psi(x,y)$ is the naive quark propagator, and $G_\chi(x,y)$ is the 
staggered quark propagator.
With this basic relationship, we can construct the heavy baryon two-point functions for $s^\pi=0^+$ using the operator $\mathcal{O}_5$.
\begin{eqnarray}
C_{5}(\vec{p},t)&=&\sum_{\vec{x}}e^{-i \vec{p} \cdot \vec{x}} \langle 
\mathcal{O}_{5} (\vec{x},t) \overline{\mathcal{O}}_{5} (\vec{0},0) \rangle \\
\label{trace}
&=&\sum_{\vec{x}}e^{-i\vec{p}\cdot\vec{x}} \epsilon_{abc} \epsilon_{a'b'c'} \mathbf{tr} [ 
G_1^{aa' T} (x,0) C \gamma_5 G_2^{bb'} (x,0) C \gamma_5 ] G_{H}^{cc'} (x,0)\\
&=&\sum_{\vec{x}}e^{-i\vec{p}\cdot\vec{x}} \epsilon_{abc} \epsilon_{a'b'c'}
4 G_{1\chi}^{aa'} (x,0) G_{2\chi}^{bb'} (x,0) G_{H }^{cc'} (x,0).
\end{eqnarray}
$G_1(x,0)$ and $G_2(x,0)$ are naive light quark propagators, $G_H(x,0)$ is the Wilson type heavy quark
propagator, and $G_{1\chi}(x,0)$ and $G_{2\chi}(x,0)$ are staggered light quark propagators.
 Note that the trace in
Eq.~\ref{trace} is for spinor indices, not for color indices. 
Similarly, we can derive the two-point function for $s^\pi=1^+$ with the operator $\mathcal{O}_\mu$, which is
\begin{equation}
C_{\mu\nu}(\vec{p},t)=\sum_{\vec{x}}e^{-i\vec{p}\cdot\vec{x}} \epsilon_{abc} \epsilon_{a'b'c'}
4 (-1)^{x_\mu} \delta_{\mu\nu} G_{1\chi}^{aa'} (x,0) G_{2\chi}^{bb'} (x,0) G_{H }^{cc'} (x,0).
\label{c_munu}
\end{equation}
We note that $\delta_{\mu\nu}$ in Eq.~\ref{c_munu} is an artifact of using
Eq.~\ref{win1} to convert the light staggered propagtor to a naive propagator.
With Wilson or clover light quarks, $C_{\mu\nu}$ would have off diagonal
elements.  We shall now elucidate the origin of the $\delta$-function.

\section{Taste mixing}
We know that heavy-light mesons do not have any taste mixing problem \cite{wingate}. 
Since we use two staggered quarks for singly heavy baryons, the taste structure of the operators might be more complicated than in the heavy-light meson case.
Furthermore, presence of the $\delta$ function in Eq.~\ref{c_munu} suggests we investigate more seriously.

We have resolved this problem by considering the taste and spin structure of the operators explicitly. 
Let's look at the operator $\mathcal{O}_\mu$ in Eq.~\ref{operator}. Since the heavy quark field $\Psi_H$ carries no taste index, we can simplify the problem by considering the di-quark operator $\mathcal{D}_\mu$ rather than the original operator $\mathcal{O}_\mu$, i.e.,
\begin{equation}
\mathcal{D}_\mu = \psi_1^{T} C \gamma_\mu \psi_2. 
\label{diquark}
\end{equation}
Next, we have a relation between the naive quark and the staggered quark fields
\begin{equation}
\psi^{\alpha'}(x)=\Omega^{\alpha' a}(x)\chi^a(x),
\label{psi_chi}
\end{equation}
where $\psi^{\alpha'}$ is the naive quark, $\chi^a$ are four copies of staggered quark, $\Omega(x)$ is the matrix given in Eq.~\ref{omega}, $\alpha'$ is the spinor index of the naive quark, and $a$ is the copy index. Also, we have
\begin{equation}
\chi^a(y+\xi)=2\Omega^{\dagger i \alpha}(\xi)q^{\alpha i,a}(y)
\label{chi_q}
\end{equation}
and
\begin{equation}
x=y+\xi,
\label{chi_q2}
\end{equation}
where $q^{\alpha i,a}$ is the staggered quark in the taste basis, $\alpha$ is the staggered spin index, $i$ is the taste index, $x$ is the coordinate index for all lattice sites, $y$ is the coordinate index for even sites only, and $\xi$ is the vector of each corner of the unit hypercube. Using Eqs.~\ref{psi_chi} and \ref{chi_q}, we can rewrite the naive quark $\psi^{\alpha'}(x)$ as a linear combination of the staggered quarks in the taste basis
\begin{equation}
\psi^{\alpha'}(x)=\Omega^{\alpha' a}(\xi)\chi^a(y+\xi)=\Omega^{\alpha' a}2\Omega^{\dagger i \alpha}(\xi)q^{\alpha i,a}(y).
\label{psi_q}
\end{equation}
Now, we can put Eq.~\ref{psi_q} into the di-quark operator, and sum over the hypercube in order to get the continuum di-quark operator $\mathcal{D}^{conti}_\mu(y)$,
\begin{eqnarray}
\label{d_conti}
\mathcal{D}^{conti}_\mu(y)&=&\sum_\xi (\psi^T_1(x)C\gamma_\mu\psi_2(x)) \nonumber \\
&=& \sum_\xi  
2 \Omega^{\dagger i \alpha}(\xi) q^{\alpha i, a}(y) 
\Omega^{T a \alpha ' }(\xi) (C\gamma_\mu)^{\alpha' \beta'} \Omega^{\beta' b}(\xi) 
2 \Omega^{\dagger j \beta}(\xi) q^{\beta j,b}(y) \\
&=&\sum_\xi  
4 \Omega^{\dagger i \alpha}(\xi) q^{\alpha i, a}(y)
(-1)^{\xi_\mu}(C \gamma_\mu)^{a b}
 \Omega^{\dagger j \beta}(\xi) q^{\beta j,b}(y), \nonumber
\end{eqnarray}
using the relation
\begin{equation}
\Omega^{T a \alpha ' }(\xi) (C\gamma_\mu)^{\alpha' \beta'} \Omega^{\beta' b}(\xi)
= (-1)^{\xi_\mu}(C \gamma_\mu)^{a b}.
\end{equation}
Next we sum over $\xi$ in Eq.~\ref{d_conti}, using
\begin{equation}
 \sum_\xi \Omega^{\dagger i \alpha}(\xi) 
(-1)^{\xi_\mu}
 \Omega^{\dagger j \beta}(\xi) = 4 (C\gamma_\mu)^{\alpha \beta} \otimes 
 (\gamma_\mu C^{-1})^{i j}.
 \label{}
 \end{equation}
Finally, the continuum di-quark operator $\mathcal{D}_{\mu }^{conti}(y)$ can be written
\begin{equation}
\mathcal{D}_{\mu }^{conti}(y) =    
16 q^{\alpha i, a}(y) (C\gamma_\mu)^{\alpha \beta} \otimes 
 (\gamma_\mu C^{-1})^{i j}
q^{\beta j,b}(y)
(C \gamma_\mu)^{a b}.
 \label{}
 \end{equation}
 From Ref.~\cite{nagata}, we immediately notice that this continuum di-quark operator $\mathcal{D}^{conti}_\mu$ overlaps with a physical $s^\pi=1^+$ spin parity state with a taste singlet. Thus, the original operator $\mathcal{O}_\mu$ overlaps with $\frac{1}{2}^+$ and $\frac{3}{2}^+$ total spin parity states of the heavy baryon without any taste mixing problem in the continuum limit. 
 
 Similarly, we can easily derive the continuum di-quark operator $\mathcal{D}^{conti}_5$ for the operator $\mathcal{O}_5$
\begin{equation}
\mathcal{D}_5^{conti}(y) = 16 q^{\alpha i, a}(y) (C\gamma_5)^{\alpha \beta} \otimes 
 (\gamma_5 C^{-1})^{i j}
q^{\beta j,b}(y)
(C \gamma_5)^{a b},
 \label{}
 \end{equation}
and we can state that the operator $\mathcal{O}_5$ overlaps with physical $\frac{1}{2}^+$ total spin parity state of the singly heavy baryon with a taste singlet for the same reason.

In addition, if we consider the two-point function $C^{conti}_{\mu\nu}$ of the continuum di-quark operator $\mathcal{D}^{conti}_\mu$, then it appears
\begin{eqnarray}
C^{conti}_{\mu\nu}(y;0)
\label{c_conti_1}
 &=& 16^2 \mathbf{Tr}[G_1(y,0) (C\gamma_\mu)\otimes(C\gamma_\mu)^\dagger
 G_2(y,0)  (C\gamma_\nu)^\dagger\otimes(C\gamma_\nu)] \nonumber \\
 &\;&\;\;\;\;\;\;\;\;\;\;\;\;\;\;\; \times (C \gamma_\mu)_{ab}(\gamma_\nu C^{-1})_{b'a'} \delta_{bb'}\delta_{aa'}  \\
 \label{c_conti_2}
  &=& 16^2 \mathbf{Tr}[G_1(y,0) (C\gamma_\mu)\otimes(C\gamma_\mu)^\dagger
   G_2(y,0)  (C\gamma_\nu)^\dagger\otimes(C\gamma_\nu)] \nonumber \\
    &\;&\;\;\;\;\;\;\;\;\;\;\;\;\;\;\; \times \mathbf{Tr}[(C \gamma_\mu)(C \gamma_\nu)^{\dagger}],
\label{}
\end{eqnarray}
where
\begin{equation}
\mathbf{Tr}[(C \gamma_\mu)(C \gamma_\nu)^{\dagger}] = 4 \delta_{\mu\nu}.
\end{equation}
As we can see from Eq.~\ref{c_conti_2}, $C^{conti}_{\mu\nu}(y;0)$ consists of two traces. 
The first trace contains spinor and taste indices. The second trace is over copy indices.
 Now, we can address where the delta function in Eq.~\ref{c_munu} came from: cancellations of the copy indices govern the delta function. Furthermore, we notice that even though the copy indices have a non-trivial structure, we can still use the operator $\mathcal{O}_\mu$, because the contribution of the copy indices is a constant, and independent of spin and taste.

\section{Results}

We use three ensembles of  $20^3\times64$ MILC coarse dynamical lattice gauge configurations 
with lattice spacing $a\approx 0.12$ fm \cite{MILC01}.
The details of the ensembles are shown in Table~\ref{table1} .
For each configuration, we require several propagators for the valence quarks. 
We compute nine different staggered light quark propagators with masses between
0.005 and 0.02, and three staggered strange quark propagators with masses 0.024, 0.03, 
and 0.0415, and each propagator is calculated with four different time sources. For the heavy  quarks, we use hopping parameter values $\kappa=0.122$ for the charm quark, and $\kappa=0.086$ for the bottom quark 
based on tuning for our heavy-light meson decay constant calculation \cite{bare} \cite{bare2}.

\begin{table}[htdp]
\begin{center}
\begin{tabular}{ccccc}
\hline
$\beta$ & $am_l$ & $am_s$ & \# of conf. for charm & \# of conf. for bottom \\
\hline
6.76 & 0.007 & 0.05 & 545 & 554 \\
6.76 & 0.01 & 0.05 & 591 & 590 \\
6.79 & 0.02 & 0.05 & 459 & 452 \\
\hline
\end{tabular}
\end{center}
\caption{Parameters of the ensembles, and number of configurations used for each heavy quark.  The light sea quark mass is $m_l$ and the strange sea quark mass is $m_s$.}
\label{table1}
\end{table}

For the data analysis for the baryon propagators, we use the same fitting and extrapolation method as in our previous work \cite{na}.

\begin{figure}[ht]
\centering
\includegraphics[width=.45\textwidth]{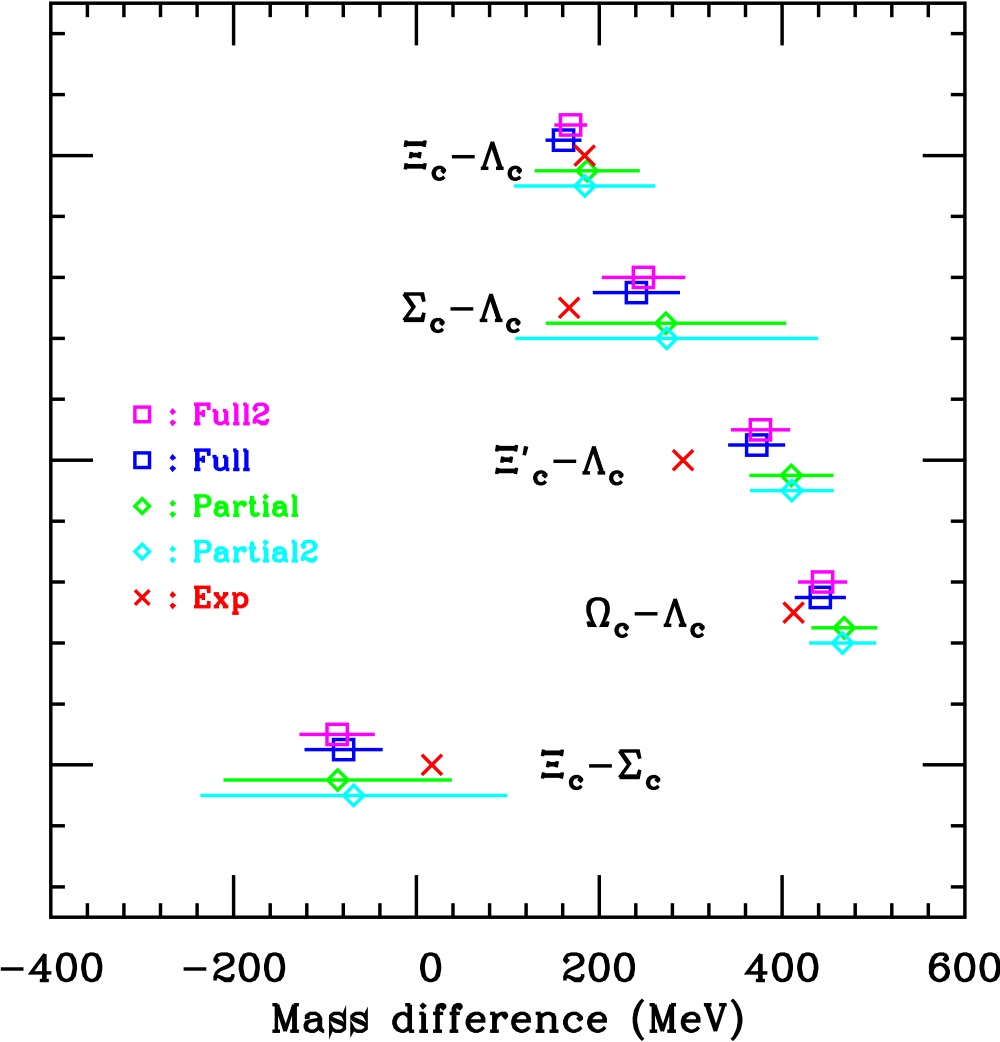}
\includegraphics[width=.45\textwidth]{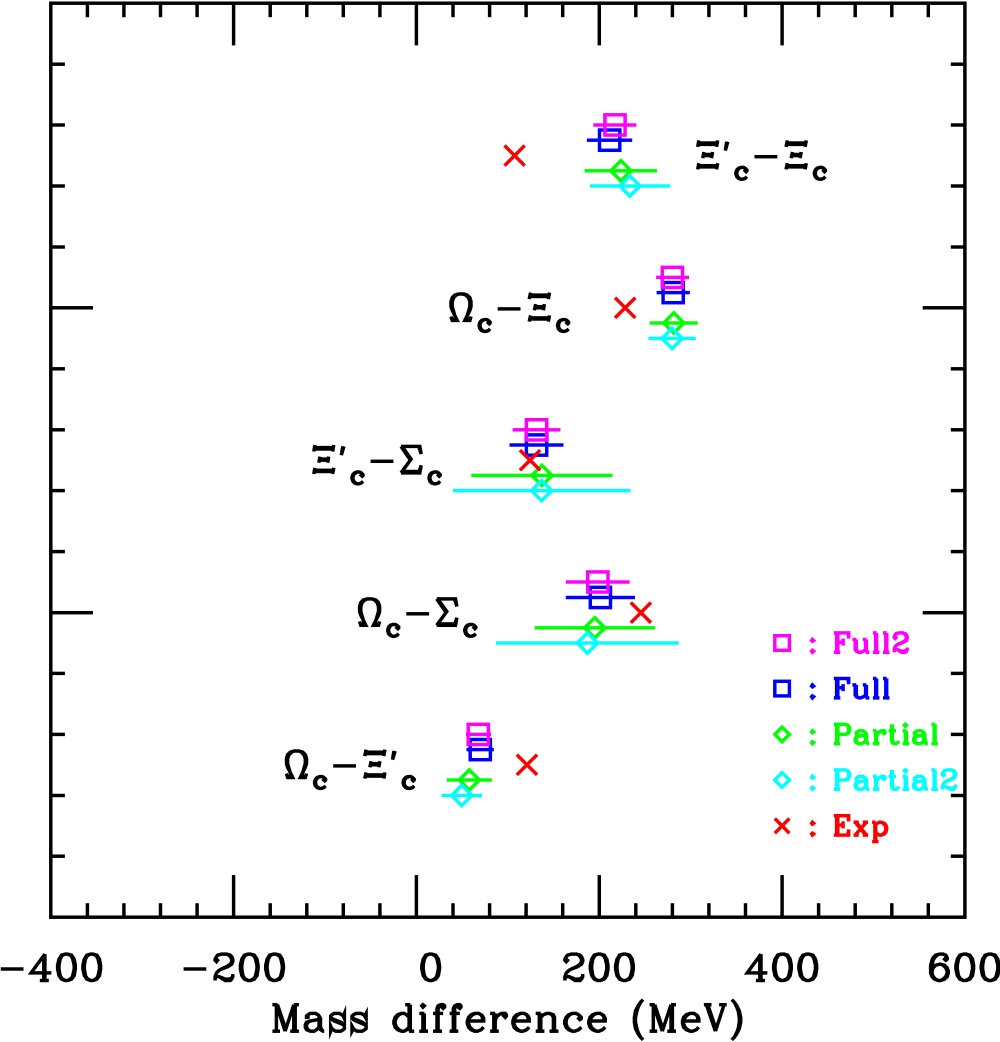}
\caption{The mass splittings for singly charmed heavy baryons.}
\label{fig1}
\end{figure}

We present mass splittings in MeV for singly charmed heavy baryons in Fig.~\ref{fig1}.
Crosses correspond to the experimental data \cite{pdg}, and squares and diamonds with error bars indicate our results. 
The experimental 
errors are considerably smaller than our own, so we ignore them here, and the errors of our results are statistical errors.
Since we do not have partially quenched staggered chiral perturbation theory for singly heavy baryons to fit both sea and valence quark mass dependences simultaneously,
we use four slightly different extrapolation methods. The result from each method is labeled as Full, Full2, Partial, or Partial2 in Fig.~\ref{fig1}. 
The first method takes all partially quenched data, and linearly extrapolates for  the valence quark masses and then the sea quark masses; this is the ``Partial'' result. 
Alternatively, we consider only data where light valence masses equal the corresponding sea quark mass and do a single extrapolation; this is the ``Full''  data. 
Further, we can obtain the mass splittings by a second method.  Instead of
applying the chiral exptrapolations to the masses themselves, we apply them to
the splittings.  Results from the second procedure are denoted with a "2".

\begin{figure}[ht]
\centering
\includegraphics[width=.45\textwidth]{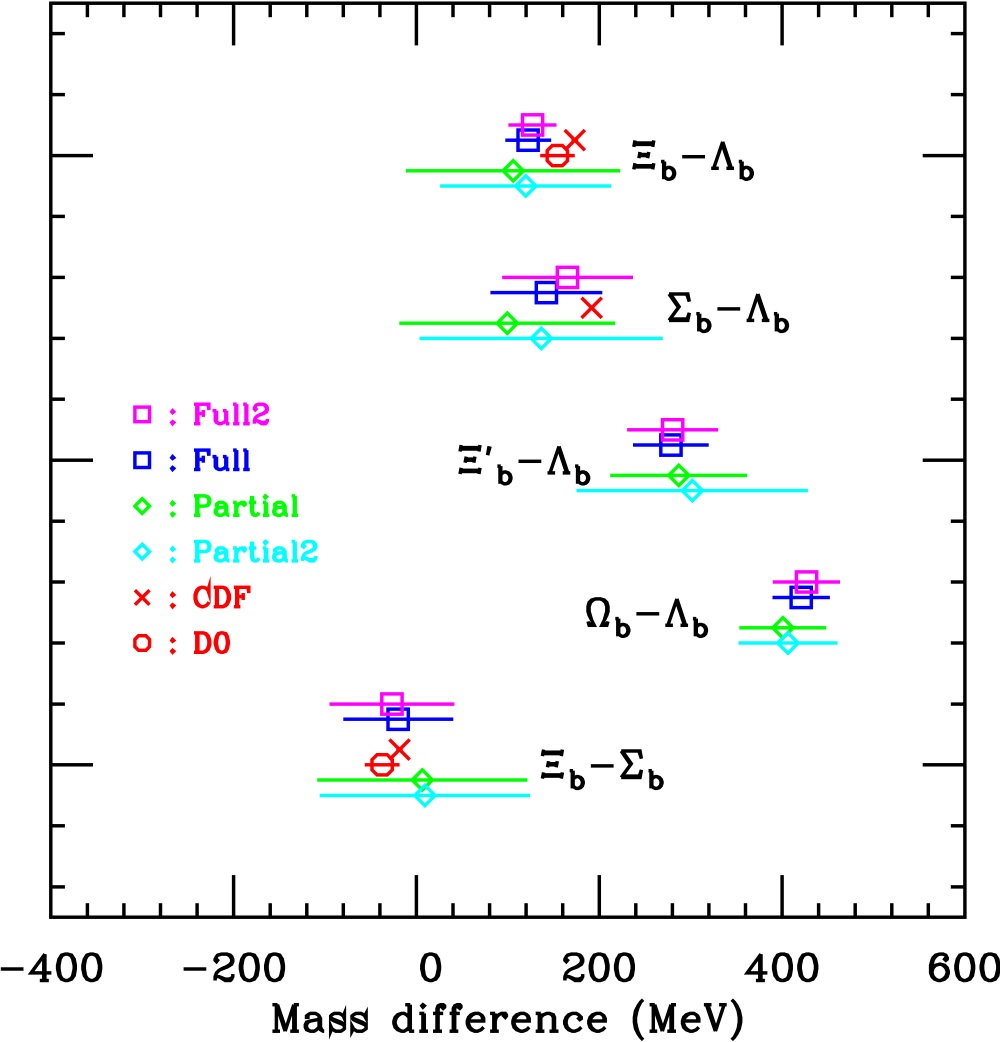}
\includegraphics[width=.45\textwidth]{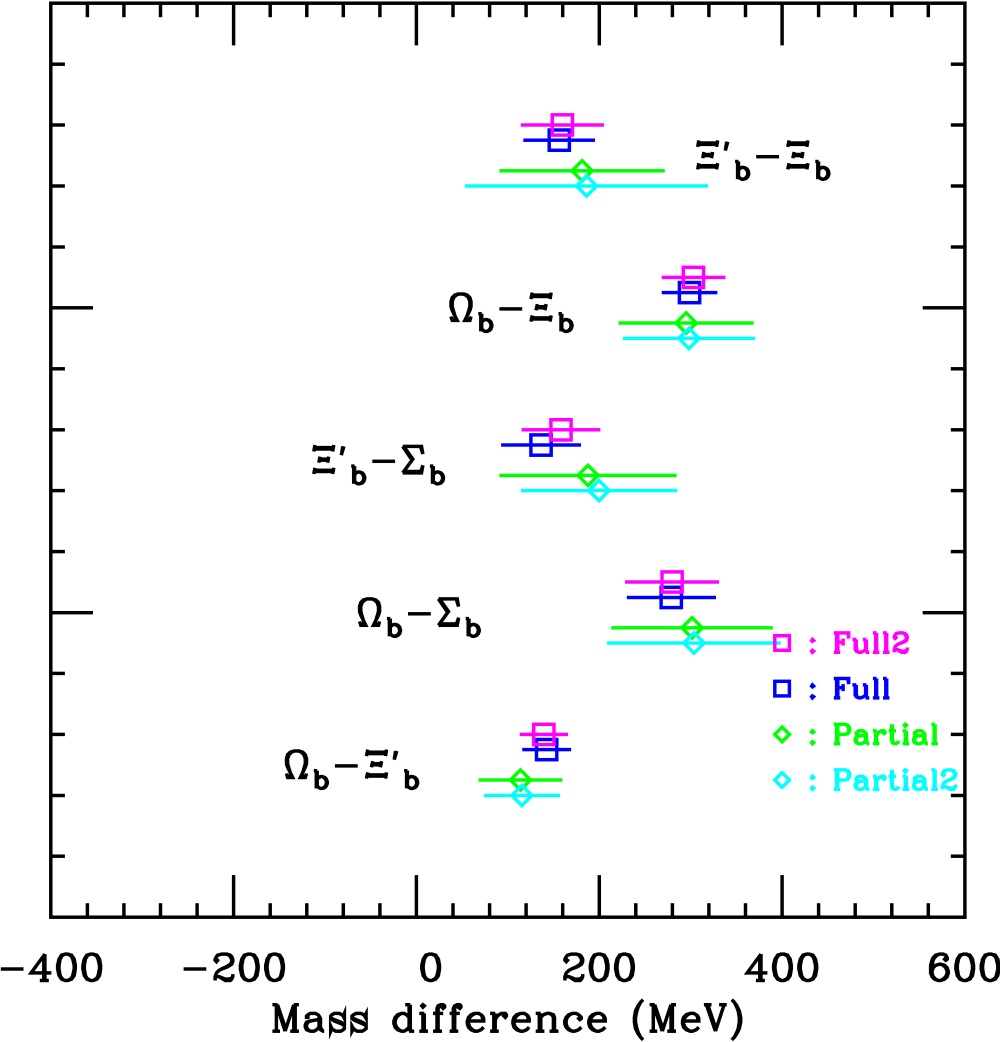}
\caption{The mass splittings for singly bottom heavy baryons.}
\label{fig2}
\end{figure}

We also present mass splittings for singly bottom heavy baryons in Fig.~\ref{fig2}.
Crosses are the recent experimental data from the CDF \cite{CDF}, and octagons are the experiment data from the D$\emptyset$ \cite{D0}. We include error bars for the D$\emptyset$ results, since the errors of the D$\emptyset$ results are larger than the other experiments, about 18 MeV. 

\begin{figure}[ht]
\centering
\includegraphics[width=.45\textwidth]{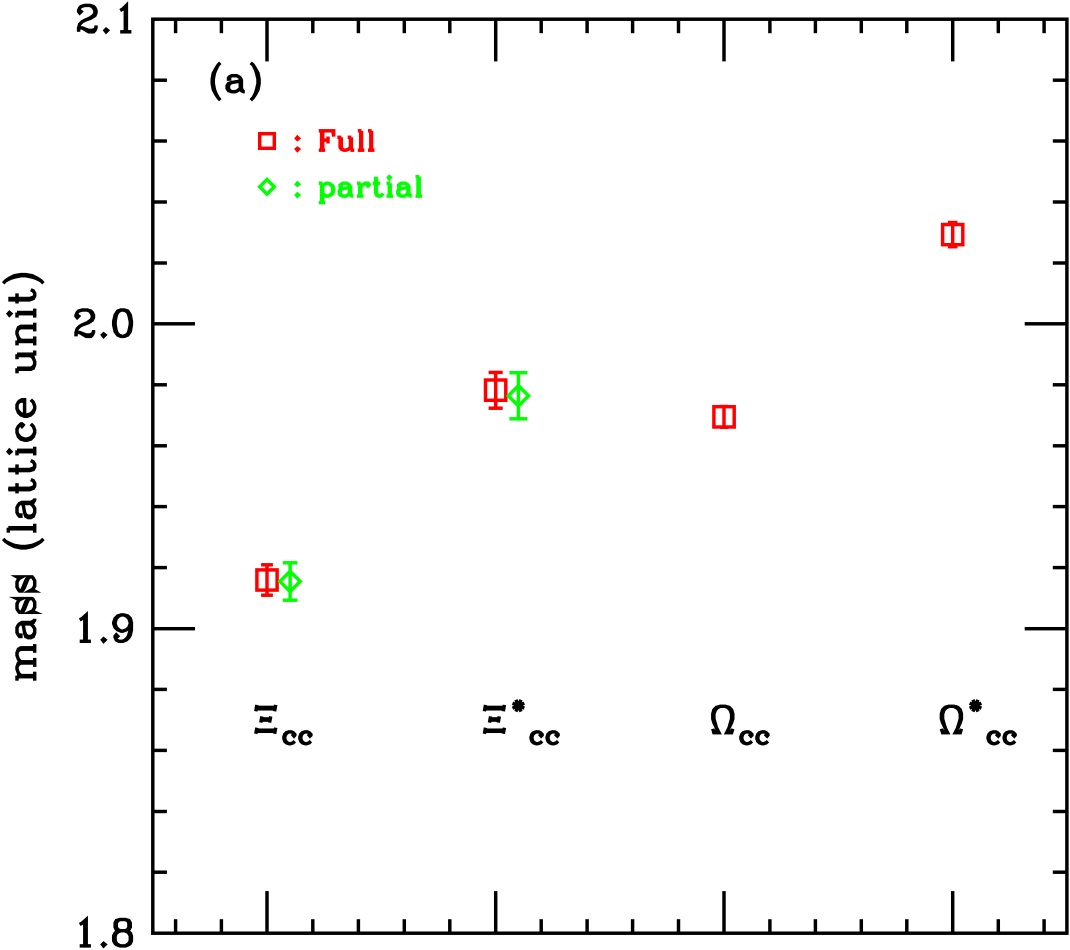}
\includegraphics[width=.45\textwidth]{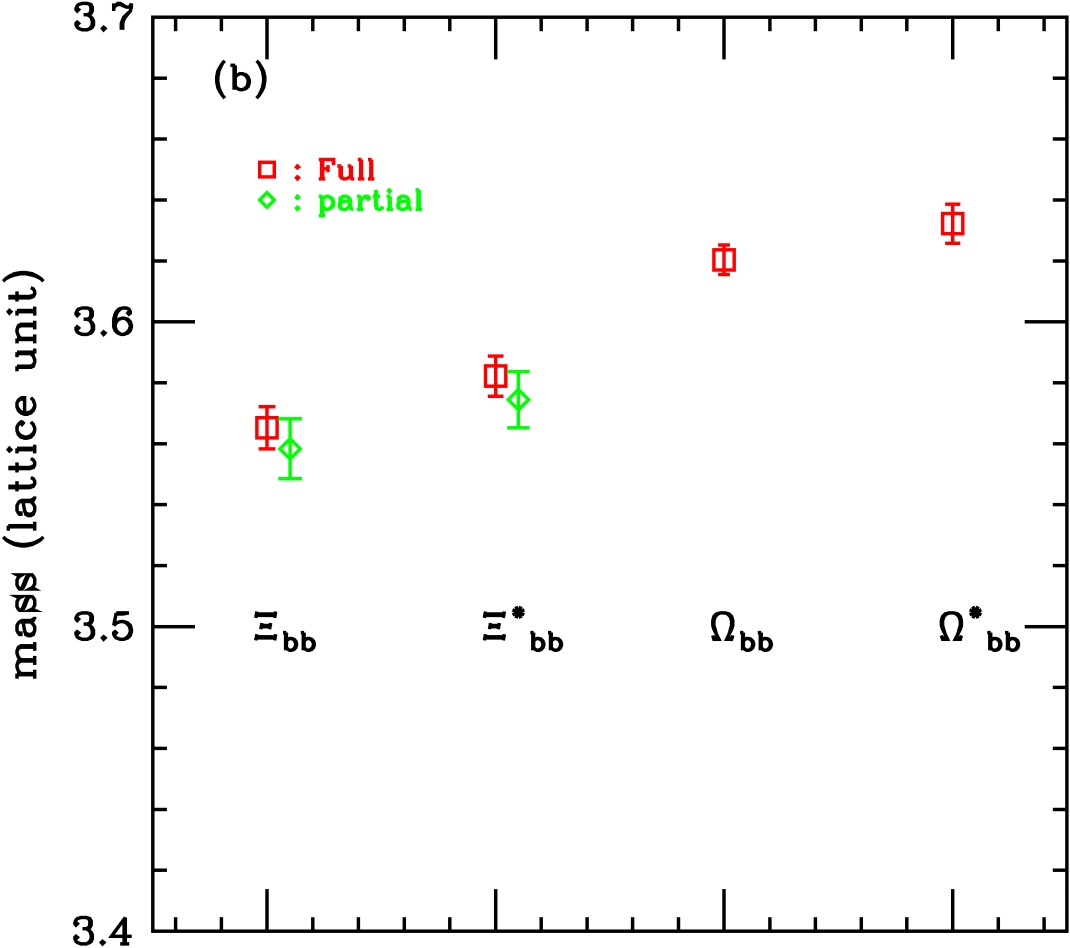}
\caption{The mass spectrum for doubly charm (a), and bottom (b) heavy baryons. }
\label{fig3}
\end{figure}

In addition, we present a preliminary mass spectrum in lattice units for doubly bottom and charmed heavy baryons in Fig.~\ref{fig3}.

\section{Future study}
We have presented the mass spectrum for charm and bottom heavy baryons based on dynamical lattice QCD, and a solution of the taste mixing problem of the operators.
The study can be extended in several ways.
In the near future, we would like to extend this work into  $28^3\times96$ MILC fine lattice gauge configurations; so that we can determine the mass spectrum more accurately. 
In this work, we apply the simple linear extrapolations for sea quark masses and valence quark masses separately; as a result, our partially quenched data analysis gives larger errors than the errors from full QCD data analysis. In order to extrapolate for the both sea and valence quark masses simultaneously and accurately, we would like to develop partially quenched staggered chiral perturbation theory for heavy baryons. Next, we must  consider systematic errors, such as finite size effects and discretization errors. In addition, we are  interested in the $\frac{3}{2}^+$ excited state masses of singly heavy baryons, which cannot be studied with our current method.
\acknowledgments{We are grateful to Claude Bernard for valuable comments on the taste mixing problem, and Kazuhiro Nagata for helpful discussions and comments. Numerical calculations were performed on the Pion cluster at Fermilab, and on the BigRed cluster at Indiana University. This work was supported in part by the U.S. DOE.}

\end{document}